\documentclass[pre,twocolumn,preprintnumbers,amsmath,amssymb,nofootinbib,floatfix]{revtex4}

\usepackage{graphicx,bm, tikz}
\usepackage[breaklinks]{hyperref}
\usepackage[normalem]{ulem}
\usepackage{comment}

\makeatletter
\def\graphicscale{\twocolumn@sw{0.3}{0.4}}
\def\graphicthreescale{\twocolumn@sw{0.3}{0.4}}

\begin{document}

\title{Out-of-equilibrium scaling of the particle density in quantum
  fermionic wires \\ after a critical quenching of the chemical
  potential}

\author{Haralambos Panagopoulos} 
\affiliation{Department of Physics, University of Cyprus,
P.O. Box 20537, 1678 Nicosia, Cyprus}

\author{Ettore Vicari} 
\affiliation{Dipartimento di Fisica dell'Universit\`a di Pisa,
        Largo Pontecorvo 3, I-56127 Pisa, Italy}

\date{\today}

\begin{abstract}
We study the out-of-equilibrium scaling behavior of the particle
density in quantum fermionic Kitaev wires, after instantaneous quantum
quenches (QQs) of the chemical potential within their quantum critical
region. The critical scaling of the ground-state particle density is
known to be subleading at its Ising-like quantum transition, hidden by
regular and logarithmic terms arising from peculiar mixings with the
identity operator. This situation changes along the out-of-equilibrium
dynamics arising from QQs of the chemical potential to the critical
point, starting from the ground state for Hamiltonian parameters
within the critical region.  We analytically show that the difference
between the post-QQ particle density and its critical value develops
an out-of-equilibrium scaling behavior, in terms of the dynamic
scaling variable $\theta\sim t/\xi^z$ (where $t>0$ is the post-QQ
time, $\xi$ is the length scale of the initial state, and $z$ is the
dynamic critical exponent) associated with the post-QQ time evolution.
The scaling function turns out to have a peculiar singular behavior in
the $\theta\to 0$ limit, apparently related to the anomalous
equilibrium scaling behavior of the particle density at the starting
point of the QQ protocol.  This provides analytical evidence of
earlier conjectures on the general emergence of post-QQ dynamic
scaling behaviors of the subtracted particle density (supported by
numerical finite-size scaling analyses), unlike their equilibrium
counterpart which turns out to be dominated by nonuniversal
contributions.

\end{abstract}

\maketitle


\section{Introduction}
\label{intro}

The behavior of the ground-state particle density in interacting
many-body systems is generally nonuniversal at quantum transitions
driven by the chemical potential~\cite{RV-21,CPV-14}, due to a
nontrivial interplay between regular (short-ranged) and nonanalytic
(critical) contributions, hiding the most interesting universal
features related to the critical modes.  This is analogous to what
happens at classical thermal transitions~\cite{WK-74, Fisher-74,
  Wegner-76,PV-02}, where the energy density is dominated by regular
terms, which can be shown to arise from a nontrivial mixing with the
identity operator in renormalization-group (RG) and field-theoretical
frameworks, while the scaling terms arising from the critical modes
are only subleading.

This problem is particularly evident at the quantum transitions of
one-dimensional fermionic models~\cite{CPV-14,RV-24}, or quantum
Kitaev wires~\cite{Kitaev-01}.  Due to their equivalence with the $XY$
spin chains, the quantum transition of Kitaev wires belongs to the
two-dimensional (2D) Ising universality class characterized by the
length-scale exponent $\nu=1$ and dynamic exponent $z=1$, see, e.g.,
Refs.~\cite{Sachdev-book,RV-21}.  At their quantum critical point the
equilibrium particle density is dominated by a standard regular term
as usual, but it also shows a subleading logarithmic term arising from
the resonance of the RG weights of RG
operators~\cite{Wegner-76,CHPV-02,CPV-14}, belonging to the {\em
  energy} and {\em identity} conformal families within a
conformal-field-theory (CFT) framework (see, e.g.,
Ref.~\cite{CHPV-02}). Therefore, the genuine scaling term featuring
the critical exponents appears as the next-to-next-to-leading term in
the asymptotic expansion that describes the approach to the critical
point.  An analogous behavior characterizes the energy density at the
thermal transitions of 2D classical Ising-like models, where a
resonance between energy and identity CFT operators gives rise to the
logarithmic divergence of the specific
heat~\cite{Wegner-76,PV-02,CHPV-02}. Other notable examples of these
behaviors at classical transitions are provided by the
three-dimensional O($N$) vector models, whose scaling part of the
energy density appears only as the next-to-next asymptotic term when
the length-scale critical exponent satisfies $\nu>2/3$, as is the  case
for any $N\ge 2$~\cite{PV-02}.  Therefore the particle density at
quantum transitions and the energy density at thermal transitions are
not considered among the optimal observables to study critical
behaviors at equilibrium.

As argued in Refs.~\cite{PV-24,RV-24}, this problematic scenario for
these observables may change, and be overcome, when looking at the
out-of-equilibrium critical dynamics.  The energy density at classical
transitions and the particle density at quantum transitions (more
precisely, their differences with the critical point values) have been
conjectured to develop nontrivial dynamic scaling behaviors after
instantaneously quenching the temperature or the chemical potential
from equilibrium states around criticality to their critical points.
Indeed, it was argued~\cite{PV-24,RV-24} that their time dependences
along the post-quench critical relaxational and unitary flow have a
nontrivial out-of-equilibrium scaling limit, characterized by power
laws related to the standard critical exponents, such as the
length-scale critical exponent $\nu$, and the dynamic critical
exponent $z$ related to the classical or quantum dynamics.  In
particular, the post-quench out-of-equilibrium scaling behavior turns
out to be described in terms of the dynamic scaling variable
$\theta\sim t/\xi^z$ (where $t>0$ is the post-quench time, $\xi$ is
the correlation length of the critical modes in the initial state, and
$z$ is the dynamic critical exponent). It was also noted that the
anomalous equilibrium scaling, which characterizes the starting point
of the quench protocol, reflects itself in peculiar singular behaviors
of the out-of-equilibrium scaling functions in the small-$\theta$
limit.

These out-of-equilibrium scaling scenarios have been supported by
numerical analyses of quench protocols within finite-size scaling
(FSS) frameworks in various classical and quantum
systems~\cite{PV-24,RV-24,BPV-25-rel,BPV-25-zn}, including fermionic
Kitaev wires or quantum Ising chains.  We also mention that the
improved out-of-equilibrium scaling behavior of the energy density at
classical continuous transitions has been successfully exploited to
obtain accurate estimates of the relaxational dynamic exponent $z$ at
topological transitions in lattice gauge models~\cite{BPV-25} where
the energy density turns out to be the optimal gauge-invariant
quantity to be computed and
analyzed~\cite{BPV-25-rel,BPV-25-zn,ACP-26}.

However, establishing these conjectures on firmer grounds calls for
exact analytical results in some of the most paradigmatic models.
This paper is meant to fill this gap. For this purpose, we again focus
on the out-of-equilibrium scaling behavior of the particle density of
fermionic Kitaev wires after quenching the chemical potential within
its Ising critical region, which can be considered as a notable
theoretical laboratory. We present analytical results showing the
out-of-equilibrium scaling behavior of the particle density in the
thermodynamic limit, arising from a quantum quench (QQ) of the
chemical potential within the critical region, for example starting
from the ground state for Hamiltonian parameters in the neighbourhood
of the critical point.  For this purpose, we exploit earlier exact
expressions for the time dependence of the transverse magnetization in
quantum $XY$ chains after QQs of the transverse magnetic field,
obtained by Barouch, McCoy and Dresden~\cite{BMD-70}, which can be
straightforwadly mapped into the particle-density evolution of the
fermionic Kitaev wire.  We prove that this post-QQ evolution has a
nontrivial out-of-equilibrium scaling limit, and compute the
corresponding dynamic scaling function.  In particular, we confirm
that the out--of-equilibrium scaling describing the post-QQ time
dependence is characterized by a dynamic scaling function that is
singular in the vanishing limit of its time scaling variable $\theta$.
This analysis substantially strengthens the evidence in favor of the
conjectured out-of-equilibrium scaling behavior of the particle
density at quantum transitions driven by the chemical potential.

The paper is organized as follows. In Sec.~\ref{quacri} we present the
one-dimensional fermionic Kitaev model, and summarize the main
features of its quantum critical behavior driven by the chemical
potential, in particular that of the particle density. In
Sec.~\ref{outeq} we describe the QQ protocol to the critical point,
which gives rise to the out-of-equilibrium behavior we are interested
in; moreover we briefly report the known results for the post-QQ time
dependence of the particle density~\cite{BMD-70}, which we use to
determine its out-of-equilibrium scaling limit.  In
Sec.~\ref{outthlim} we analytically derive the out-of-equilibrium
scaling behavior of the particle density, and discuss its main
features.  In Sec.~\ref{outthlim2} we consider more general QQ
protocols, showing that out-of-equilibrium scaling behaviors for the
particle density emerge also in these cases. Finally, in
Sec.~\ref{conclu} we summarize and draw our conclusions.

\section{Particle density in fermionic Kitaev wires}
\label{quacri}

\subsection{The model and its quantum criticality}
\label{model}

The one-dimensional fermionic Kitaev model~\cite{Kitaev-01} is a
paradigmatic many-body system undergoing a continuous
zero-temperature quantum transition driven by the chemical potential.
It is defined by the Hamiltonian
\begin{equation}
  \hat H(\mu,\gamma) = - J \sum_{x} \big( \hat c_x^\dagger \hat
  c_{x+1}^{\phantom\dagger} + \gamma \hat c_x^\dagger \hat
  c_{x+1}^\dagger+{\rm h.c.}  \big) - \mu \hat{N},
  \label{kitaev}
\end{equation}
where $\hat c_x$ is the fermionic annihilation operator associated
with the $x$th site of the chain, and 
\begin{equation}
  \hat{N} = \sum_x \hat{n}_x, \qquad \hat n_x=\hat c_x^\dagger \hat
  c_x^{\phantom\dagger},
  \label{partnumb}
\end{equation}
is the particle-number operator.  The Hamiltonian parameter $\mu$
denotes the chemical potential, while $\gamma>0$ controls the relative
strength of the terms which do not conserve the fermionic number.  In
the following, we fix the energy scale by assuming $J=1$ and also set
the Planck and Boltzmann constants $\hslash = k_B = 1$.

The Hamiltonian~\eqref{kitaev} can be mapped into the diagonal
quadratic form~\cite{LSM-61, Katsura-62, Pfeuty-70, BG-85}
\begin{equation}
  \hat H(\mu,\gamma) = \sum_k E(k) \bigl(\hat a^\dagger_k \hat a_k -
  1/2\bigr),
  \label{H-harmonic}
\end{equation}
where $\hat a_k$ are related fermionic annihilation operators, which
are obtained through a suitable linear transformation of the original
$\hat c_x$ operators, and
\begin{equation}
  E(k) = \sqrt{ (\mu+2\cos k)^2 + 4\gamma^2 \sin^2k }
  \label{Ek-XY}
\end{equation}
is their dispersion relation.

By means of a nonlocal Jordan-Wigner transformation (see, e.g.,
Ref.~\cite{Sachdev-book}), the Hamiltonian~\eqref{kitaev} can be also
mapped into the Hamiltonian of the quantum $XY$ chain,
\begin{equation}
  \hat H_{XY} \!= \!- \! \sum_{x} \! \left[ \frac{1 + \gamma}{2} \hat
    \sigma^{(1)}_x \hat \sigma^{(1)}_{x+1} + \frac{1 - \gamma}{2} \hat
    \sigma^{(2)}_x \hat \sigma^{(2)}_{x+1} + g \hat
    \sigma^{(3)}_x\right]\! ,
  \label{XYchain}
\end{equation}
where $\hat \sigma^{(k)}_x$ are the Pauli matrices ($k=1,2,3$), and
\begin{equation}
g=-\mu/2,\qquad   \hat\sigma^{(3)}_x = 1 - 2 \hat{n}_x.
  \label{sign}
\end{equation}

The quantum Kitaev wires and $XY$ chains undergo 
continuous quantum transitions belonging to the same 2D Ising universality
class (see, e.g., Refs.~\cite{Sachdev-book, RV-21}), respectively at
$\mu = \mu_c = -2$ and at $g = g_c = 1$, independently of the parameter
$\gamma>0$.  We define the deviation of the relevant parameter $\mu$
from its critical value as
\begin{equation}
  w \equiv {\mu_c - \mu \over 2} = g - g_c , \qquad (\mu_c=-2, \; g_c=1).
  \label{wdef}
\end{equation}
The continuous Ising-like transition of Kitaev wires at $\mu_c$ are
characterized by the length-scale critical exponent $\nu=1$, related
to the RG dimension $y_w = 1/\nu=1$ of the relevant Hamiltonian
parameter $w$, so that the length scale $\xi$ of the critical quantum
fluctuations diverges as $\xi \sim |w|^{-\nu}=|w|^{-1}$ when
approaching the critical point $w\to 0$ at zero temperature.  The
dynamic exponent $z=1$ is associated with the unitary quantum dynamics
within this universality class. It also determines the power law
$\Delta\sim\xi^{-z}$ of the vanishing gap with increasing $\xi$.  The
temperature $T$ represents a further relevant RG perturbation at the
quantum critical point; indeed, at the critical point $w=0$ and
nonzero temperature $T$ the length scale is finite and increases as
$\xi \sim T^{-1/z}$ with decreasing $T$. Finally, we report the RG
dimension $y_n$ of the particle-density operator $\hat n_x$,
i.e.,~\cite{Sachdev-book, RV-21}
\begin{equation}
y_n = d+z-y_w = 2 - y_w = 1,
  \label{ynexp}
\end{equation}
which controls the power
laws associated with its expectation value and correlations.

The above universal critical exponents determine the asymptotic power
laws of the observables as a function of the temperature $T$ and the
chemical potential $\mu$. The asymptotic critical expansions show also
the presence of logarithmic terms~\cite{Wegner-76, SS-00, Queiroz-00,
  Salas-01, ONGP-01, IH-02, CHPV-02, IH-09, CGNP-11, Izmailian-12,
  Izmailian-13}. They can be explained by the presence of
resonances~\cite{Wegner-76} between the {\em identity} operator of RG
dimension 2 and the {\em energy} operator of RG dimension 1 within the
corresponding 2D conformal field theory (CFT) with central charge
$c=1/2$. In particular, such a resonance mechanism gives rise to the
leading logarithmic divergence of the specific heat at the 2D Ising
critical point.

\subsection{Equilibrium behavior of the particle density}
\label{equisca}

The equilibrium behavior of the particle density within the critical
region can be obtained from the derivative of the free-energy density
\begin{equation}
  F(w,T,\gamma) = - {T\over L}
  \ln \big[ {\rm Tr}\,e^{-\beta \hat H(\mu,\gamma)} \big],
  \qquad \beta = 1/T, \label{freeF}
\end{equation}
with respect to $\mu$.  The free-energy density in the thermodynamic
limit can be written as~\cite{Katsura-62}
\begin{equation}
  F(w,T,\gamma) = - \int_{0}^{\pi} \, {dk\over 2\pi}
\, \left\{ E(k) + 2T \ln\Bigl[1 +
  e^{-\beta E(k)}\Bigr]\right\},
\label{fwtga}
\end{equation}
where $E(k)$ is given in Eq.~\eqref{Ek-XY}.  In the critical limit
$w\to 0$ and $T\to 0$, it behaves as (see Refs.~\cite{CPV-14,RV-24}
for more details)
\begin{eqnarray}
&&  F(w,T,\gamma) \approx F_{\rm reg}(w,\gamma)
  + \label{FscalXY}\\ && \quad
  + {\sqrt{\gamma^2 + w} \over 4\pi} u_w^2 \ln u_w^2
    - {2 \sqrt{\gamma^2 + w} \over \pi}  u_t^2 \,f(u_w/u_t). \nonumber
\end{eqnarray}
$F_{\rm reg}(w,\gamma)$ is a regular function at the critical point,
which is independent of $T$~\cite{CPV-14, RV-21}, and can be expanded
in powers of $w$; $u_w$ and $u_t$ are the scaling fields corresponding
to the relevant parameters $w$ and $T$, which are given by
\begin{eqnarray}
u_w =
{w\over \sqrt{\gamma^2 + w}},\qquad u_t = {T\over 2\sqrt{\gamma^2 + w}};
\label{uwut}
\end{eqnarray}
finally,
\begin{equation}
  f(x) = \int_0^\infty dz \, \ln \Bigl(1 + e^{-\sqrt{x^2 + z^2}}\Bigr) 
  \label{fxsca}
\end{equation}
is a universal scaling function (apart from a factor and the
normalization of the argument).

The equilibrium particle density $\varrho_e$ is obtained by
differentiating $F(w,T,\gamma)$ with respect to $\mu$.  Assuming
translational invariance, and keeping only the most relevant terms,
\begin{eqnarray}
  &&\varrho_e \equiv {\rm Tr} \,[\rho_G(\beta) \, \hat{n}_x] 
  = -\frac{\partial F}{\partial\mu} = \frac{1}{2} \frac{\partial F}{\partial w}
  \label{rhobeh}\\ &&\;\;\approx \varrho_{\rm reg}(w,\gamma)+
        {1\over 2\pi} \left( u_w \ln |u_w| + u_w\right) - {1 \over \pi}
        u_t\, f_\rho(u_w/u_t), \nonumber
\end{eqnarray}
where $\rho_G(\beta) = e^{-\beta\hat{H}}/{\rm Tr} [e^{-\beta \hat H}]$
is the Gibbs density matrix, and $f_\rho(x)=\partial_x f(x)$. In the
critical limit the particle density is dominated by the contribution
of the regular term, which can be expanded as
\begin{equation}
  \varrho_{\rm reg}(w,\gamma) = a_0(\gamma) + a_1(\gamma)\, w + ...
  \label{rhoreg}
\end{equation}
where $a_i$ are nonuniversal constants depending on $\gamma$.  In
particular, $a_0=1/2 - 1/\pi$ for $\gamma=1$.

Note that also the behavior of the difference
\begin{equation}
  D_e(w,T,\gamma) \equiv \varrho_e(w,T,\gamma) - \varrho_c(\gamma),
  \quad \varrho_c(\gamma)=a_0(\gamma),
  \label{deltarhodef}
\end{equation}
is dominated by the nonuniversal logarithmic term arising from the
resonance between identity and energy operators, hiding the universal
scaling behavior of the critical modes,~\cite{CPV-14, RV-21}
\begin{equation}
  \varrho_{\rm scal} = T^{y_n/z} f_\rho(u_w/u_t) \approx w \,
  \widetilde{f}_\rho(T/w),
  \label{rhosca}
\end{equation}
which remains logarithmically suppressed with respect to the leading
term (the scaling functions $f_\rho$ and $\widetilde{f}_\rho$ are
straightforwardly related). As we shall see, this problematic behavior
disappears when considering the out-of-equilibrium dynamics arising
from QQs.

We finally mention that the above scaling behaviors can be extended to
finite-size systems, within FSS frameworks, see, e.g.,
Refs.~\cite{CPV-14, RV-21, FB-72, Barber-83, Privman-90, Cardy-editor,
  CHPV-02, PV-02}.

\section{Post-quench dynamics of the particle density}
\label{outeq}

\subsection{Quantum quench to the critical point}
\label{quprot}

We want to study the out-of-equilibrium quantum dynamics of the
particle density arising from instantaneous QQs of the parameter
related to the chemical potential, from $\mu\neq \mu_c$ to the critical
point $\mu_c$, or correspondingly from $w\equiv (\mu_c-\mu)/2\neq 0$
to $w_c=0$. We only consider {\em soft} quenches starting from initial
conditions close to the critical point (i.e., for small values of
$|w|$ and temperature $T$), so that the system stays always within the
critical regime during the post-QQ quantum evolution.  Therefore, we
focus on the following QQ protocol:

(i) At $t=0$ the system is prepared in an equilibrium state $\rho_i$
of the Hamiltonian $\hat H(w,\gamma)$, cf. Eq.~\eqref{kitaev}, for a
given value of $w\neq 0$, which may be the Gibbs state at a finite
temperature, i.e.,
\begin{equation}
\rho_i= \rho_G(\beta)={e^{-\beta\hat{H}(w,\gamma)}\over {\rm Tr}
  [e^{-\beta \hat H(w,\gamma)}]},
\label{rhoigibbs}
\end{equation}
or the ground state $|\Psi_0(w)\rangle$, i.e.,
\begin{equation}
  \rho_i = |\Psi_0(w)\rangle \langle \Psi_0(w)|,
\label{rhoiground}
\end{equation}
corresponding to the $\beta\to \infty$ limit in the absence of
degeneracy.

(ii) At $t>0$, the system evolves unitarily driven by the critical
Hamiltonian $\hat H(w_c,\gamma)$, i.e.,
\begin{equation}
{d \rho(t) \over dt}  = - i [\hat H(w_c,\gamma), \rho(t)], 
\qquad \rho(t=0)=\rho_i.
  \label{scheq}
\end{equation}
Notice that the driving dynamics is assumed to be always unitary,
i.e., that of a closed quantum system, even in the case we start from
a finite-temperature Gibbs distribution. This means that the external
bath is only used to prepare the initial finite-temperature Gibbs
state, but it is assumed to be disconnected from the system for $t>0$.

(iii) The out-of-equilibrium post-QQ dynamics of the particle density
is monitored by computing its expectation value
\begin{equation}
  \varrho(t,w,\gamma) = {\rm Tr}\,[\rho(t)\,\hat{n}_x],
  \label{varrhotdef}
\end{equation}
where we assumed translational invariance. The relation with the
corresponding transverse magnetization of the equivalent quantum $XY$
chain can be easily derived using Eq.~(\ref{sign}).

In the following we mostly consider QQ protocols starting from
ground states, and  focus on the time dependence of the subtracted
particle density
\begin{equation}
  D(t,w,\gamma) = \varrho(t,w,\gamma) - \varrho_c(\gamma),
  \label{sigmasdef}
\end{equation}
where $\varrho_c(\gamma)$ is the equilibrium value of the particle
density at the critical point $\mu_c=-2$.

\subsection{Post-quench evolution of the particle density}
\label{postQQ}

The post-QQ time dependence of the particle density can be
straightforwardly obtained from the solutions of the corresponding
problem within the quantum $XY$ model reported in Ref.~\cite{BMD-70}.
This allows us to derive an exact expression for the time dependence
of the particle density $\varrho$ after a quench from $\mu_i$ to
$\mu_c=-2$ with $\mu_i = \mu_c - 2 w$.  We write the difference
$D(t,w,\gamma)$ defined in Eq.~(\ref{sigmasdef}) as a sum of two terms:
\begin{equation}
 D(t,w,\gamma) = X(w,\gamma) + Y(t,w,\gamma),
  \label{faf}
\end{equation}
where the first term $X$ provides the asymptotic stationary behavior,
while $Y$ is a time dependent term vanishing in the large-time limit.
Defining
\begin{equation}
  Q(k,w,\gamma) = \sqrt{(1 - \cos k+w)^2 + \gamma^2 \sin^2k},
  \label{lagk}
\end{equation}
the functions entering Eq.~(\ref{faf}) are
\begin{eqnarray}
  X(w,\gamma) & = & Z(w,\gamma) - Z(0,\gamma),\label{fagg0s}\\
  Z(w,\gamma) & = & \frac{1}{2} - \frac{1}{2\pi}\int_0^\pi dk\,
\frac{1-\cos k}{Q(k,w,\gamma)
  Q(k,0,\gamma)^2}
\quad\label{fagg0}\\ 
  & \times &
  \left[(1-\cos k+w)(1-\cos k) +\gamma^2
  \sin^2 k\right], \nonumber
\end{eqnarray}
where we used the fact that $\varrho_c(\gamma) = Z(0,\gamma)$, and
\begin{equation}
  Y(t,w,\gamma) = - {w \gamma^2\over 2\pi} \int_0^\pi dk \,
  {\sin^2k \,\cos[4Q(k,0,\gamma)t] \over Q(k,w,\gamma)
    Q(k,0,\gamma)^2}. \label{ftgg0}
\end{equation}
The large-time limit of these formulas, keeping $w$
fixed, can be also derived from the results reported in
Ref.~\cite{BMD-70}.  For example, for $\gamma=1$,
\begin{eqnarray}
  D(t,w,1)= X(w,1)-{w \sin(8 t - \pi/4) \over 32\sqrt{\pi}
    (2+w)t^{3/2}}
+ O(t^{-5/2}),\quad
    \label{fr1infty}
\end{eqnarray}
where 
\begin{eqnarray}
X(w,1) \approx - w^2\left(\frac{\ln w}{8\pi} + a \right)\quad
  \label{pw1d}
\end{eqnarray}
for $w\ll 1$, with $a = - 0.023055246396$.

These expressions provide the starting point of our analysis. In the
following section we show that they admit an out-of-equilibrium
scaling limit, which leads to a more standard scaling behavior of the
subtracted particle density, controlled by the critical exponents
$\nu$ and $z$, as conjectured in earlier works~\cite{PV-24,RV-24}.
Some interesting features of the above exact
expressions have been discussed in Refs.~\cite{PEF-12,RV-20,RV-21}.

\section{Out-of-equilibrium scaling}
\label{outthlim}

We now derive an out-of-equilibrium scaling ansatz for the post-QQ
dynamics of the subtracted particle density in the thermodynamic
limit. On the basis of earlier theoretical works on dynamic critical
phenomena, see, e.g.,
Refs.~\cite{HH-77,Ma-book,Dziarmaga-05,CEGS-12,PRV-18,TV-22,RV-21,RV-24},
we identify the ratio $t/\xi^z \sim t \,w^{z\nu}$ as the relevant
scaling variable associated with time.  Therefore we introduce the
dynamic scaling variable
\begin{equation}
  \theta = w^{z\nu} t = w \, t.
\label{thetadef}
\end{equation}
The out-of-equilibrium scaling behavior of the particle density was
discussed within a FSS framework in Ref.~\cite{RV-24}, leading to the
hypothesis of the emergence of an out-of-equilibrium FSS, which was
supported by numerical computations in the FSS limit.  Then, by
performing an appropriate infinite-size limit, the out-of-equilibrium
scaling ansatz 
\begin{equation}
  D(t,w,\gamma) \approx {w^{y_n\nu}\over \gamma} \, \Gamma(\theta),
  \qquad y_n\nu =1 ,
  \label{dtscalinf}
\end{equation}
was put forward in the thermodynamic limit, where $y_n$ is the RG
dimension of the particle-density operator, cf. Eq.~(\ref{ynexp}),
$\Gamma(\theta)$ is a scaling function that is supposed to be
universal apart from a multiplicative factor and the normalization of
the argument, which may depend on $\gamma$.  Actually, as we shall
see, the prefactor of this ansatz and the definition of the scaling
variable $\theta$ are chosen so that $\Gamma(\theta)$ turns out to be
completely independent of the Hamiltonian parameter $\gamma$.

\begin{figure}[!t]
  \includegraphics*[scale=\graphicscale]{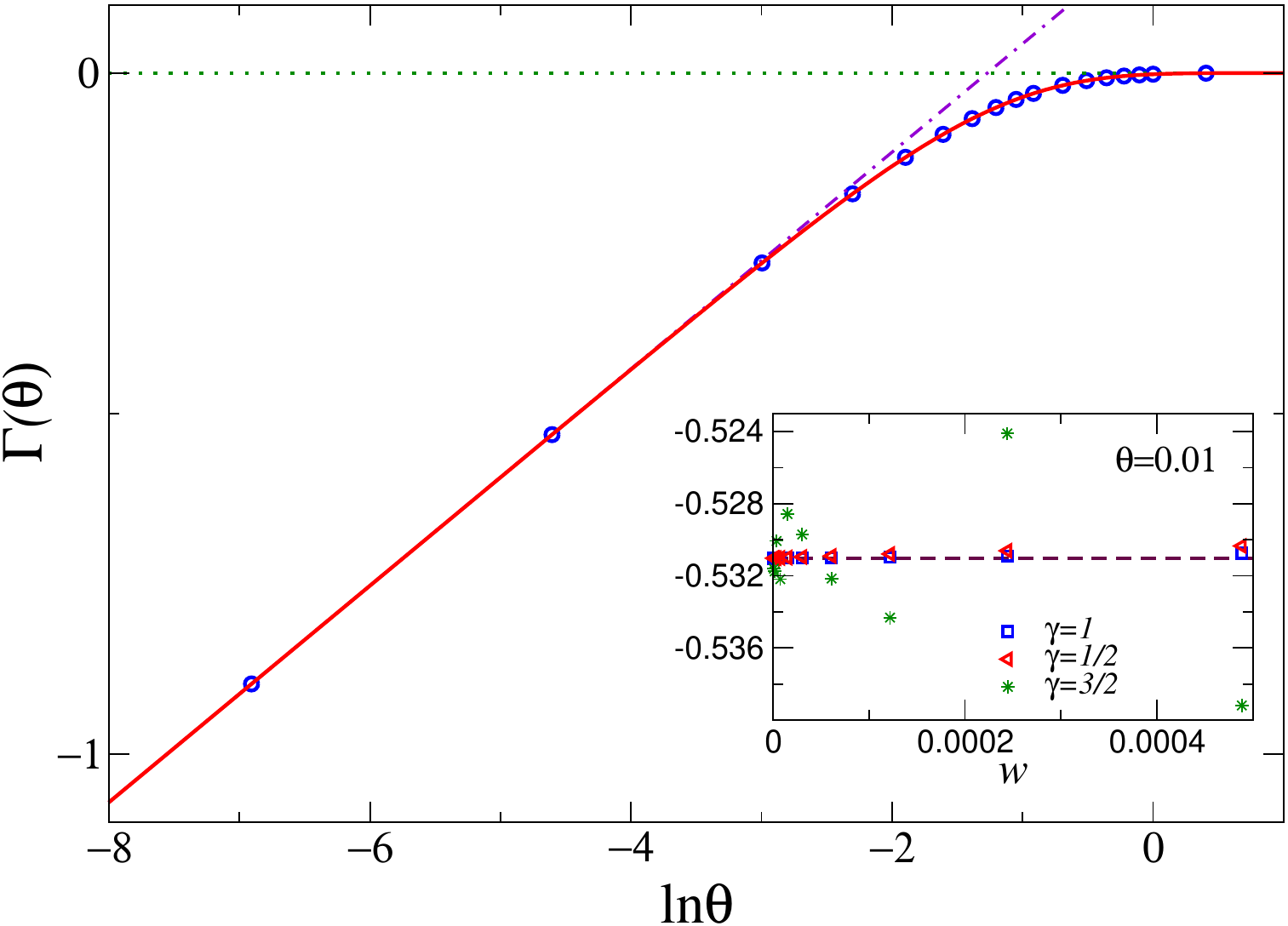}
  \caption{ We show the scaling function $\Gamma(\theta)$ (full line)
    obtained in Eq.~(\ref{dinfyexa}), and compare it with the
    numerical results obtained by taking the out-of-equilibrium
    scaling limit, i.e., $t\to\infty$ and $w\to 0$ keeping $\theta$
    fixed, of $\gamma D(t,w,\gamma)/w$ (whose estimated errors are
    very small, indeed they are almost invisible in the figure).  The
    dot-dashed and dotted lines show the asymptotic behaviors for
    $\theta\to 0$ and $\theta\to\infty$, respectively.  The inset
    shows the approach to the asymptotic value $\Gamma(\theta)$ for
    $\theta=0.01$ and various values of $\gamma$, i.e.,
    $\gamma=1,\,1/2,\,3/2$. The dashed line indicates the asymptotic
    value $\Gamma(0.01)=-0.531027...$. }
      \label{scalingfunction}
\end{figure}

We now show that the post-QQ time dependence of the subtracted
particle density $D(t,w,\gamma)$ behaves as in Eq.~(\ref{dtscalinf})
in the limit $t\to \infty$ and $w\to 0$ keeping $\theta=wt$ fixed, and
analytically derive the scaling function $\Gamma(\theta)$.  We first
note that only the time-dependent integral $Y(t,w,\gamma)$ contributes
to the asymptotic scaling function $\Gamma(\theta)$, essentially
because the time-independent term $X(w,\gamma)$ behaves as
\begin{equation}
  \frac{\gamma X(w,\gamma)}{w} \approx - {w\ln w\over 8\pi \gamma^2}
\quad{\rm for}\;\;w\to 0,  
\label{asydepw}
\end{equation}
thus it vanishes for $w\to 0$. To derive the dynamic scaling function
$\Gamma(\theta)$, we also note that the integral (\ref{ftgg0}) in the
dynamic scaling limit entailing $t\to\infty$ is essentially determined
by the small values of $k$ (this fact can be easily confirmed
numerically). Therefore, the asymptotic scaling limit does not change
if we approximate and replace
\begin{eqnarray}
\sin k \approx k, \qquad Q(k,0,\gamma)\approx \gamma k,
\label{sinkk}
\end{eqnarray}
and
\begin{equation}
Q(k,w,\gamma) \approx w \sqrt{1+q^2}, \qquad q = \gamma k/w,
  \label{lagka1}
\end{equation}
in the integrand of $Y(t,w,\gamma)$, so that
\begin{equation}
  Y(t,w,\gamma) \approx  - {w \over 2\pi\gamma} \int_0^{\gamma\pi/w} dq \,
  {\cos(4 q \theta) \over \sqrt{1+q^2}}, \label{ftgg0a1}
\end{equation}
after a variable change from $k$ to $q$.  In the dynamic scaling limit
$w\to 0$ keeping $\theta=wt$ fixed, we can also replace $\gamma
\pi/w\to\infty$ in the upper bound of the integral, obtaining the
final expression
\begin{eqnarray}
  \Gamma(\theta) = -\frac{1}{2\pi} \int_0^\infty
  dq\,{\cos(4q\theta)\over \sqrt{1+q^2}} = 
  - \frac{1}{2\pi} K_0(4\theta),
  \label{dinfyexa}
\end{eqnarray}
where $K_0(x)$ is a modified Bessel function of the second kind.  This
result confirms that $\Gamma(\theta)$ is universal, i.e., it is
independent of $\gamma$.

The scaling function $\Gamma(\theta)$ shows an unusual singular
behavior for $\theta\to 0$.  Indeed, its asymptotic expansion for
$\theta\to 0$, see, e.g., Ref.~\cite{math}, shows a logarithmic
divergence:
\begin{eqnarray}
  \Gamma(\theta) = {1\over 2\pi} \left(\ln \theta + \ln 2 +
  \gamma_E\right) + O(\theta^2\ln\theta).
  \label{smalltheta}
\end{eqnarray}
Note that this logarithmic behavior, and in particular its
coefficient, turns out to match the leading logarithmic term of the
equilibrium subtracted particle density, cf. Eq.~(\ref{rhobeh}),
recalling that $\theta=wt$.  We also note that the convergence to the
out-of-equilibrium scaling limit $\Gamma(\theta)$ cannot be uniform
for $\theta\to 0$.  On the other hand, for large $\theta$ we have a
regular exponential approach to zero:
\begin{eqnarray}
  \Gamma(\theta) = - \frac{1}{\sqrt{32\pi\theta}} e^{-4\theta}
  \left[ 1 + O(\theta^{-1})\right].
  \label{largetheta}
\end{eqnarray}

In Fig.~\ref{scalingfunction} we show the scaling function
$\Gamma(\theta)$. To further support its validity, we compare it with
numerical integrations of $D(t,w,\gamma)$ in the out-of-equilibrium
scaling limit, i.e., by extrapolating the numerical data of $\gamma
D(t,w,\gamma)/w$ in the large-$t$ limit keeping $\theta=wt$ fixed.  We
have checked the independence of $\gamma$ of such extrapolations by
repeating the computations for various values of $\gamma>0$, see the
inset of Fig.~\ref{scalingfunction}.  These computations allow us to
obtain quite accurate results (the most accurate extrapolations are
generally obtained for $\gamma\le 1$), in particular for small values
of $\theta$, which are in perfect agreement with the exact solution
(\ref{dinfyexa}).  Some results on the approach to the asymptotic
scaling function $\Gamma(\theta)$ are shown in the inset of
Fig.~\ref{scalingfunction} for a fixed value $\theta=0.01$ and various
values of $\gamma$, i.e. $\gamma=1/2,\,1,\,3/2$, and several values of
$w$ in the limit $w\to 0$.  We note that the data approach the
asymptotic value $\Gamma(\theta=0.01)$ for $\gamma\le 1$, with
corrections that appear to decay approximately as $O(w\ln w)$, and
allow us to obtain accurate determinations of the scaling limit. On
the other hand, larger scaling corrections are observed for
$\gamma>1$, which are characterized by oscillations around the
asymptotic value, whose size gets suppressed in the $w\to 0$ limit.

Finally, we mention that the dynamic scaling limit can be extended to
allow for a finite temperature related to the Gibbs distribution of
the initial state, cf. Eq.~(\ref{rhoigibbs}). Since in the critical
regime $T\sim \xi^{-z}$ where $\xi \sim w^{-\nu}$ and the critical
exponents are $z=1$ and $\nu=1$, the scaling variable corresponding to
the temperature must be
\begin{equation}
  \tau = T/w = (\beta w)^{-1},
  \label{taudef}
\end{equation}
which is the scaling variable of the function $\widetilde{f}_\rho$ in
Eq.~(\ref{rhosca}).  Therefore we expect
\begin{equation}
  D(t,w,\gamma,T) = \varrho(t,w,\gamma,T) - \varrho_c(\gamma,T)
  \approx \frac{w}{\gamma} \,\Xi(\theta,\tau)
\label{dtwgatau}
\end{equation}
in the limit $t\to \infty$, $w\to 0$ and $T\to 0$ keeping $\theta$ and
$\tau$ fixed. Of course, $\Xi(\theta,0) = \Gamma(\theta)$.  Again we
can derive an exact expression for the dynamic scaling function
$\Xi(\theta,\tau)$, extending the above results for $T=0$. Using 
the results of Ref.~\cite{BMD-70}, to keep into account a finite
temperature the integrand of the integral $Y$ must be multiplied by
\begin{equation}
  \tanh[\beta Q(k,w,\gamma)] \approx \tanh(\sqrt{1+q^2}/\tau),
  \label{tanhtau}
\end{equation}
where $q=\gamma k/w$.  Therefore, proceeding as in the
zero-temperature case, we obtain
\begin{eqnarray}
  \Xi(\theta,\tau) = -\frac{1}{2\pi} \int_0^\infty
  dq\,
  {\tanh(\sqrt{1+q^2}/\tau)\,\cos(4q\theta)\over \sqrt{1+q^2}} .
  \label{dinfyexatau}
\end{eqnarray}

We finally mention that the post-QQ scaling behavior of the particle
density was analyzed within a FSS framework in Ref.~\cite{RV-24},
providing numerical evidence of the out-of-equilibrium scaling arising
from QQs to the critical point. The exact results presented here
should place this scenario on a much firmer ground.

\section{More general QQ protocols}
\label{outthlim2}

We now consider more general QQ protocols, in which the system starts
from the ground state associated with an initial value $w$, then it
evolves unitarily for $t>0$ driven by the Hamiltonian with a generic
value $w_e<w$, which may be also negative, and it is generally
different from $w_c=0$. In the preceding section we focused on the
particular case $w_e=w_c$. To maintain the system within the critical
regime, both $w$ and $w_e$ must be sufficiently close to the critical
point $w_c=0$.  The post-QQ dynamics for generic $w$ and $w_e$ can be
also obtained from the solutions reported in Ref.~\cite{BMD-70}.

We are interested in the subtracted particle density along the QQ
protocol defined as
\begin{equation}
  E(t,w,w_e,\gamma) = \varrho(t,w,w_e,\gamma) - \varrho_e(w_e,\gamma),
  \label{sigmasdefe}
\end{equation}
where $\varrho_e(w_e,\gamma)$ is the equilibrium value of the particle
density at $w_e$. Similarly to Sec.~\ref{outthlim}, we write it as a
sum of two terms,
\begin{equation}
 E(t,w,w_e\gamma) = {\cal X}(w,w_e,\gamma) + {\cal Y}(t,w,w_e,\gamma),
  \label{fafe}
\end{equation}
where the first term ${\cal X}$ provides the asymptotic stationary
behavior, while ${\cal Y}$ is a time dependent term vanishing in the
large-time limit.  Using the results of Ref.~\cite{BMD-70}, we write
them as
\begin{eqnarray}
&&{\cal X}(w,w_e,\gamma) = {\cal Z}(w,w_e,\gamma) - {\cal
    Z}(w_e,w_e,\gamma),\label{fagg0se}\\ &&{\cal Z}(w,w_e,\gamma) = -
  \frac{1}{2\pi}\int_0^\pi dk\, \frac{1+w_e-\cos k}{Q(k,w,\gamma)
    Q(k,w_e,\gamma)^2}   \nonumber\\
  &&\quad \times
  \left[(1-\cos k+w_e)(1-\cos k+w) +\gamma^2 \sin^2 k\right],
  \qquad\label{fagg0ee}
\end{eqnarray}
where we used $\varrho_e(w_e,\gamma) = {\cal
  Z}(w_e,w_e,\gamma)$, and
\begin{eqnarray}
&& {\cal Y}(t,w,w_e,\gamma) = - {(w-w_e)\gamma^2\over 2\pi} \times
\nonumber \\
&&\quad \times \int_0^\pi dk \,
  {\cos[4Q(k,w_e,\gamma)t] \sin^2k \over Q(k,w,\gamma)
    Q(k,w_e,\gamma)^2}.
  \label{ftgg0ee}
\end{eqnarray}

\begin{figure}[!t]
  \includegraphics*[scale=\graphicscale]{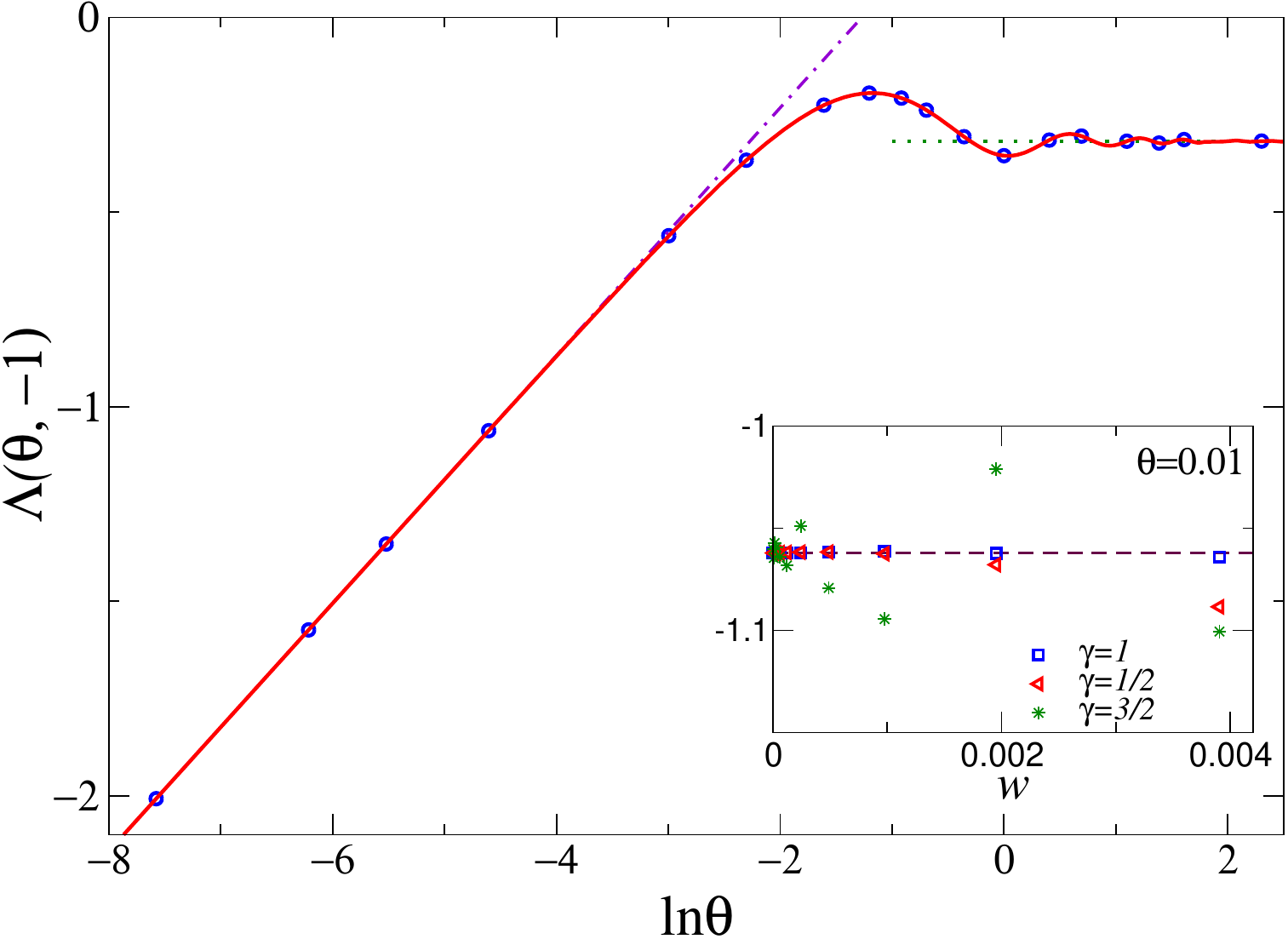}
  \caption{ We show the exact expression (\ref{dinfyexawm1gpt}) (full
    line) for the scaling function $\Lambda(\theta,u=-1)$, comparing
    it with the numerical results obtained by taking the scaling limit
    $t\to\infty$ and $w\to 0$ keeping $\theta$ and $u=w_e/w$ fixed.
    The dot-dashed and dotted lines show the asymptotic behaviors for
    $\theta\to 0$ and $\theta\to\infty$, respectively.  The inset
    shows the approach to the asymptotic value $\Lambda(\theta,-1)$
    for $\theta=0.01$ and $\gamma=1,\,1/2,\,3/2$. The dashed line
    indicates the asymptotic value $\Lambda(0.01,-1)=-1.0621814...$. }
      \label{scalingfunctionwem1}
\end{figure}

We are interested in a scaling limit analogous to that of the QQ to
the critical point, see Sec.~\ref{outthlim}, i.e.,
the limits $w,\,w_e \to 0$, $t\to \infty$ keeping $\theta=wt$
and 
\begin{equation}
  u = w_e/w
  \label{ura}
\end{equation}
fixed.  We show that the post-QQ evolution of the subtracted particle
density $E(t,w,w_e,\gamma)$ satisfies the out-of-equilibrium scaling
behavior
\begin{equation}
  E(t,w,w_e=u\,w,\gamma) \approx {w\over \gamma} \Lambda(\theta,u)
\label{dtwga}
\end{equation}
in the limit $t\to \infty$ and $w\to 0$, keeping $\theta=wt$ and
$u=w_e/w$ fixed.  Of course, the function $\Gamma(\theta)$ of the
previous section corresponds to $\Lambda(\theta,0)$.  The dynamic
scaling function $\Lambda(\theta,u)$ can be obtained by proceeding
analogously to the case $u=0$.  We obtain
\begin{eqnarray}
\Lambda(\theta,u) = A(u) + B(\theta,u),
  \label{dinfyexawu}
\end{eqnarray}
where
\begin{eqnarray}
  A(u) &=& \lim_{w\to 0} \gamma {{\cal X}(w,uw,\gamma)
    \over w}
  \label{ainfty} \\
&=& - \frac{u}{2\pi} \left[
   \ln |u| + {\rm sign}(u) {(1-u) {\rm arccos}|u|\over \sqrt{
       1-u^2}}\right],
\nonumber
\end{eqnarray}
in particular
$A(0)=0$ and $A(-1)=-1/\pi$, and
\begin{eqnarray}
  B(\theta,u) =  \frac{u-1}{2\pi}\int_0^\infty dq\,{q^2
  \cos(4 \sqrt{u^2+q^2}\theta) \over (1+q^2)^{1/2} (u^2+q^2)}.
\label{bfunc}
\end{eqnarray}
Like QQs to the critical point, i.e., the case $u=0$, the $\theta\to
0$ limit turns out to diverge logarithmically, indeed
\begin{equation} \Lambda(\theta,u)
  \approx (1-u) {\ln \theta \over 2\pi} \quad{\rm for}\;\;\theta\to 0,
\label{smallthescae}
\end{equation}
with a $u$-dependent prefactor. We note that, unlike the case $u=0$,
for generic $u\neq 0$ the coefficient of the logarithmic divergence
does not match that of the leading logarithmic term of the initial
equilibrium subtracted particle density, cf. Eq.~(\ref{rhobeh}),
corresponding to the starting point of the QQ protocol, showing that
this should not be considered as a necessary feature of the
out-of-equilibrium scaling (actually we note that it corresponds to
the difference between the leading logarithmic terms of the
equilibrium behaviors at $w$ and $w_e=u\, w$).

As a particular example, we focus on the case $u=-1$. Using
Eqs.~(\ref{ainfty}) and (\ref{bfunc}), we obtain
\begin{eqnarray}
  \Lambda(\theta,-1) = -
  \frac{1}{\pi}\left[1 + 
  \int_0^\infty dq\,{q^2 \cos(4\sqrt{1+q^2}\theta) \over (1+q^2)^{3/2}} \right],
  \quad \label{dinfyexawm1}
\end{eqnarray}
which can be written in terms of closed-form Meijer-G
representations; indeed,
using standard notation~\cite{math},
\begin{eqnarray}
  \Lambda(\theta,-1) = - \frac{1}{\pi} 
  - \frac{\theta}{2}
G^{\,1,\,2}_{\,3,\,1}
\!\left(
\frac{1}{4\theta^2}
\;\middle|\;
\begin{matrix}
\frac{3}{2},\,\frac{3}{2},\,1\\[4pt]
0
\end{matrix}
\right).
\label{dinfyexawm1gpt}
\end{eqnarray}
The asymptotic behavior for $\theta\to 0$ (see, e.g.,
Ref.~\cite{math}) turns out to be
\begin{equation}
  \Lambda(\theta,-1) = \frac{1}{\pi}\left
  (\ln \theta + \gamma_E + \ln 2\right) +  O(\theta^2\ln \theta),
  \label{dinfwm1smalltheta}
\end{equation}
while for large $\theta$ we have
\begin{equation}
  \Lambda(\theta,-1)
  = - \frac{1}{\pi} 
  -\frac{\theta^{-3/2}}{8\sqrt{2\pi}}
  \cos\left(4\theta+\frac{3\pi}{4}\right)+O(\theta^{-5/2}).
  \label{dinfwm1largetheta}
\end{equation}
The scaling function $\Lambda(\theta,u=-1)$ obtained in
Eq.~(\ref{dinfyexawm1gpt}) is shown in Fig.~\ref{scalingfunctionwem1},
where it is also compared with numerical computations of the dynamic
scaling by extrapolations using the original Eqs.~(\ref{sigmasdefe}),
(\ref{fagg0se}), (\ref{fagg0ee}) and (\ref{ftgg0ee}), which fully
agree. As shown by the inset of Fig.~\ref{scalingfunctionwem1}, the
main features of the approach to the asymptotic scaling behavior are
analogous to those observed for $u=0$ in the previous section.  In
particular, the convergence is quite stable for $\gamma\le 1$, with
$O(w)$ corrections (apart from logarithms), analogously to the $u=0$
case.

\section{Conclusions}
\label{conclu}

We investigate the out-of-equilibrium scaling behaviors of the
particle density at continuous quantum transitions driven by the
chemical potential in many-body systems without particle number
conservation.  We challenge earlier hypotheses~\cite{RV-24} on the
asymptotic scaling properties of the particle density along time
evolutions driven by instantaneous QQs of the chemical
potential within the critical region, which are conjectured to emerge
even though the equilibrium scaling behavior is generally hidden by
leading terms related to nontrivial mixings with identity operator,
such as regular terms. For this purpose, we focus on the fermionic
Kitaev model (equivalent to quantum $XY$ chain), which provides a
notable theoretical laboratory of interacting particle systems where
to check scaling scenarios at continuous quantum transitions.

We present analytical results for the out-of-equilibrium scaling
behavior of the particle density, arising from a QQ of the chemical
potential within the critical region. In particular, we consider QQ
protocols starting from the ground state for Hamiltonian parameters
within the critical region.  We prove the emergence of a nontrivial
out-of-equilibrium scaling behavior for the subtracted particle
densities (\ref{sigmasdef}) and (\ref{sigmasdefe}) in terms of the
time scaling variable $\theta$, cf. Eq.~(\ref{thetadef}).  We report
analytical expressions for the out-of-equilibrium scaling functions in
the thermodynamic limit, obtained by taking an appropriate
out-of-equilibrium scaling limit of the known post-QQ time evolutions
reported in Ref.~\cite{BMD-70}. The dominant nonuniversal logarithmic
term appearing in the equilibrium behavior, see Eq.~(\ref{rhobeh}),
is removed along the post-QQ dynamics. However, the scaling
functions are characterized by peculiar logarithmic singular behaviors
when $\theta\to 0$, which reflects the anomalous equilibrium scaling
behavior at the starting point of the QQ protocol.  Analogous results
are obtained for more general QQ protocols within the critical region.

Earlier evidence in favor of the conjectured out-of-equilibrium
scaling behavior of the particle density was essentially based on
numerical FSS analyses~\cite{RV-24}. Therefore the analytical results
presented in this paper substantially strengthen the validity of this
scenario within quantum many-body systems at criticality.

Analogous out-of-equilibrium scaling behaviors are generally expected
at any quantum transition, when considering observables related to the
derivative of the free-energy density with respect to the relevant
Hamiltonian parameter preserving the symmetry that drives the quantum
transition, such as the transverse magnetization at the quantum
transitions of $d$-dimensional quantum Ising systems, or the square
angular momentum at the quantum transitions of $d$-dimensional quantum
rotor models~\cite{Sachdev-book,RV-21}.

It is worth mentioning that there are theoretical
proposals~\cite{SD-12, PD-23}, as well as experimental attempts to
realize fermionic Kitaev wires, by means of quantum
dots~\cite{Dvir-etal-23}, integrated circuits~\cite{IYMHYE-23}, or
even quantum computers~\cite{Huang-etal-21, SBEP-21, Rancic-22,
  Mi_etal-22}.  In this context, observables related to the particle
density should be accessible, therefore there could be the possibility
to study its out-of-equilibrium behavior, achieving a further
effective probe of the universal features at quantum transitions
(unlike its equilibrium behavior, which is essentially dominated by
the nonuniversal short-range fluctuations).

We also remark that the exact results presented here at quantum
transitions also provide an indirect confirmation of analogous
conjectures for the out-of-equilibrium scaling behaviors of the energy
density at classical thermal transitions. They were already exploited
to achieve accurate determinations of the dynamic exponent $z$ at
topological transitions of lattice gauge systems where the energy
density turns out to be the optimal gauge-invariant
observable~\cite{BPV-25-rel,BPV-25-zn,ACP-26}.  It is also worth
mentioning that the critical relaxational flow at finite-temperature
transitions is similar to the so-called gradient flow that is often
exploited to numerically study the four-dimensional lattice quantum
chromodynamics (QCD), which is the theory of strong
interactions~\cite{Weinberg-book,Creutz-book}, in order to define a
running coupling from the lattice energy density, see, e.g.,
Refs.~\cite{Luscher-10,LW-11}.  Indeed, since the continuum limit of
lattice QCD is realized in the zero-coupling (zero-temperature) limit,
where the lattice length scale diverges exponentially, the critical
(fixed-point) relaxational flow becomes a simple deterministic
gradient flow at vanishing bare gauge coupling (corresponding to zero
temperature), which is equivalent to a Langevin equation without
stochastic term~\cite{Ma-book}.  On the basis of perturbative
analyses~\cite{Luscher-10,LW-11,HN-16}, the energy density along the
gradient flow is expected to develop an out-of-equilibrium scaling
behavior leading to a scaling running coupling in the continuum
limit. Analogous behaviors have been put forward for the
two-dimensional $N$-vector models with $N\ge 3$~\cite{MS-15,MSS-15},
whose critical behavior occurs in the zero-temperature limit like
lattice QCD. Therefore, the particle density along the critical
unitary flow of quantum many-body models and the energy density along
critical relaxational and gradient flows share the same improved
out-of-equilibrium scaling behavior with respect to that at
equilibrium.

These properties of the post-QQ out-of-equilibrium behavior of the
particle density at quantum transitions and the energy density at
thermal transitions can be exploited to probe their critical
behaviors, for example to achieve accurate determinations of the
critical exponents, because they can be easily computed, and in some
cases represent the optimal observable to look at, such as the case of
topological transitions in lattice gauge theory~\cite{BPV-25} where no
local order parameters exist at their continuous
transitions~\cite{BPV-25-rel,BPV-25-zn,BV-26}.

\end{document}